\documentclass[conference]{IEEEtran}
\usepackage{subfigure}
\usepackage{amsmath,amssymb,amsfonts}
\usepackage{algorithm}
\usepackage[noend]{algpseudocode}

\usepackage{array}
\usepackage{textcomp}
\usepackage{stfloats}
\usepackage{url}
\usepackage{verbatim}
\usepackage{graphicx}
\usepackage{cite}
\usepackage[dvipsnames]{xcolor}

\usepackage{hyperref}
\hypersetup{
    colorlinks=true,
    linkcolor=blue,
    filecolor=black,      
    urlcolor=blue,
    citecolor=blue,
}
\usepackage[top=0.7in, bottom=0.7in, left=0.7in, right=0.7in]{geometry}

\begin{document}

\title{Multi-Agent DRL for QoS and Energy Optimization in RIS-Enabled Open-RAN Industrial 6G TN/NTN Networks}
\author{
    \IEEEauthorblockN{Marwan Dhuheir, Thang X. Vu, and Symeon Chatzinotas,
    }
    \IEEEauthorblockA{The Interdisciplinary Centre for Security, Reliability and Trust (SnT), University of
Luxembourg, Luxembourg.
   }
}
\maketitle

\begin{abstract}
Industrial 6G networks require ultra-reliable, low-latency, and energy-efficient connectivity in dynamic and blockage-prone environments, where conventional terrestrial deployments often fail to ensure stable coverage. Hence, in this paper, we propose a RIS-enabled Open-RAN framework for integrated terrestrial/non-terrestrial (TN/NTN) industrial 6G networks, in which UAVs-mounted reconfigurable intelligent surfaces (RISs) cooperate with ground radio units and a high-altitude platform (HAP) to enhance connectivity for dense industrial IoT devices. Owing to the high dimensionality and strong coupling among decision variables, conventional optimization techniques become computationally intractable. To overcome this limitation, the joint optimization problem of data rates, latency, and energy consumptions is formulated as a decentralized partially observable Markov decision process (Dec-POMDP) and solved using a multi-agent deep reinforcement learning framework. Simulation results show improvements of up to 75\% in data rate, 25\% latency reduction, and 16\% energy savings compared with state-of-the-art learning-based and non-RIS baselines, demonstrating the effectiveness of RIS-assisted Open-RAN intelligence for industrial 6G networks.
\end{abstract}


\section{Introduction}

The rapid digitalization towards industrial systems of 6G networks is driving unprecedented yet critical requirements, including deterministic latency, extra ultra-high reliability, and sustainable energy consumption to support mission-critical industrial IoT (IIoT) applications \cite{10529728}. Large-scale factories, logistics hubs, and smart production sites are characterized by dense device deployments, dynamic obstacles, and harsh propagation environments, which render traditional cellular infrastructures insufficient for meeting the stringent performance targets envisioned for 6G networks \cite{9732420}. Consequently, new architectural paradigms capable of adapting to spatiotemporal variations and heterogeneous service demands are essential. Recent advances in open radio access network (Open-RAN) architectures provide a promising foundation for such adaptive systems by enabling disaggregated control across radio units, distributed units, and centralized intelligence layers \cite{10528242}. By exposing standardized interfaces and supporting AI-native control through RAN intelligent controllers (RICs), Open-RAN allows flexible and intelligent resource orchestration and real-time network reconfiguration \cite{nguyen2024emerging}. 

To further enhance coverage and resilience in industrial environments, non-terrestrial network components such as unmanned aerial vehicles (UAVs) and high-altitude platforms (HAPs) are increasingly considered as integral elements of 6G systems \cite{10622205,she2023guest}. UAVs, in particular, offer flexible positioning and rapid deployment, making them well suited for overcoming blockage and coverage holes\cite{11161898}. When equipped with reconfigurable intelligent surfaces (RISs), UAVs can passively manipulate the wireless propagation environment, providing energy-efficient signal enhancement without active transmission c\cite{11062573}. However, the managing the resource allocations and RIS phase shift elements in such heterogenous environments, coupled with spectrum sharing among terrestrial and aerial links, introduces strong interdependencies between communication, control, and energy management that challenge conventional optimization approaches.

Most existing work addresses these challenges using centralized optimization or heuristic-based solutions, which often fail to scale and lack robustness in highly dynamic industrial scenarios \cite{10396846,11007537}. Although learning-based methods have been explored for RIS configurations and resource allocation, they typically ignore RIS-enabled propagation control, TN/NTN integration, or Open-RAN-based intelligence\cite{11150413,bhuyan2024distributed,10628007}. Moreover, the absence of multi-timescale coordination mechanisms limits their applicability to practical industrial deployments.

Motivated by these limitations, this paper introduces a multi-agent deep reinforcement learning (MADRL) framework for RIS-enabled Open-RAN industrial 6G networks with integrated terrestrial and non-terrestrial layers. In the proposed system, UAV-mounted RISs cooperate with ground radio units (GRUs) and a HAP to provide reliable and energy-efficient connectivity to distributed IIoT devices. The joint problem of computing resources, RIS phase configuration, spectrum sharing, and resource allocation is modeled as a decentralized partially observable Markov decision process (Dec-POMDP). An efficient learning structure is adopted to provide an efficient solution for the formulated problem of the joint optimization of throughout of the data collection, execution and transmission latency, and energy consumption to process the industrial data frames.

The main contributions of this work can be summarized as follows:
\begin{itemize}
\item We develop a RIS-enabled Open-RAN architecture for integrated TN/NTN industrial 6G networks, leveraging UAV-mounted RISs to enhance coverage and reliability in challenging industrial environments.
\item We formulate a comprehensive optimization problem that captures computing resources, RIS configuration, spectrum sharing, and QoS provisioning under practical latency and energy constraints.
\item We propose a MADRL framework that enables scalable, adaptive, and decentralized control of the learned policies.
\item Extensive simulations validate the proposed approach, demonstrating notable gains in data rate by 75\%, latency reduction by 25\%, and energy consumption by 16\% over state-of-the-art baselines, including multi-agent actor-critic (MAAC), multi-agent proximal policy optimization (PPO) without RIS inclusion and centralized-based baseline.
\end{itemize}

The rest of this article is organized as follows: Section \ref{info_system_model} presents the description of our system model. In Section \ref{problem_formulation}, we delineate the problem formulation. Section \ref{Performance_evaluation} explains the implementation results of the proposed approach. At the end, section \ref{conclusion} concludes and discusses future research directions.

\section{system model}
\label{info_system_model}
\begin{figure}[t]
    \centering
\includegraphics[width=0.30\textwidth]{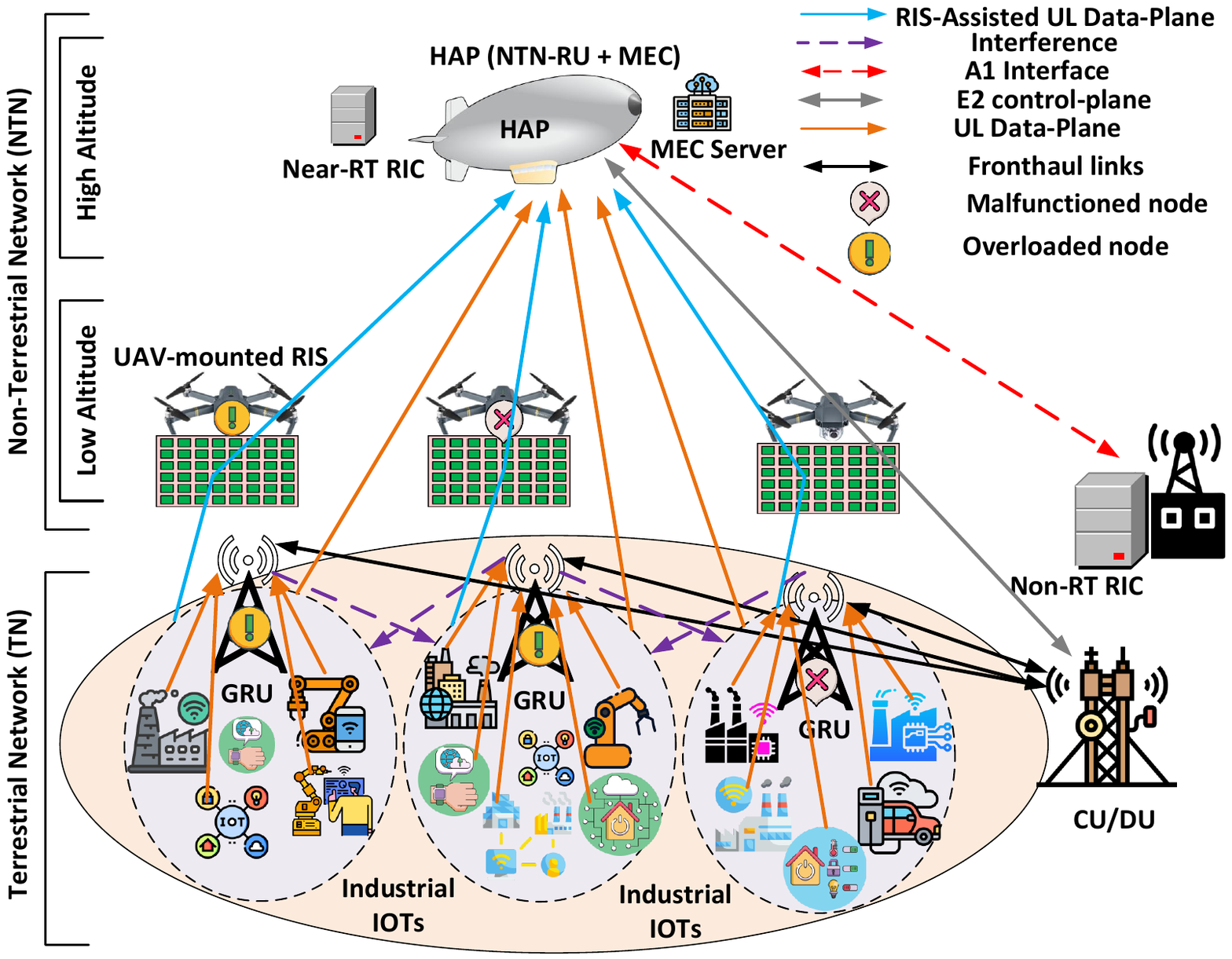}
    \caption{System model.}
    \label{fig:System_Model}
\end{figure}

Fig.~\ref{fig:System_Model} illustrates multi UAVs-mounted RISs-assisted Open-RAN TN/NTN architecture tailored for uplink-centric industrial communications. The system consists of a HAP, GRUs, UAV-mounted RISs, and distributed IIoT devices, where IIoT nodes transmit data such as sensor measurements, machine status, and alarm signals. Due to severe blockage, traffic bursts, and infrastructure impairments in industrial environments, terrestrial GRUs alone may fail to ensure reliable connectivity. To overcome these limitations, UAVs-mounted RISs passively enhanced by reconfiguring the wireless propagation environment and assisting congested GRU links. This model is well suited for industrial applications including smart manufacturing, predictive maintenance, robotic monitoring, and post-disaster operations, where reliable and low-latency data collection is critical. We also assume that the HAP is equipped with a co-located MEC server and executes offloaded tasks locally, while CU/DU entities are responsible only for RAN control and coordination.

Both GRUs and UAV-mounted RISs serve the set of IIoT devices $i \in \mathcal{I}=\{1,2,\ldots,I\}$ distributed over the coverage area. Let $u \in \mathcal{U}=\{1,2,\ldots,U\}$ and $g \in \mathcal{G}=\{1,2,\ldots,G\}$ denote the sets of UAV-mounted RISs and GRUs, respectively. The IIoT devices produce set of tasks of $j \in \mathcal{J} = \{1,\cdots,J\}$ to be executed either locally or offload them to one of the connected nodes. The service area is partitioned into disjoint clusters, where every cluster is assisted by at most one UAV-mounted RIS. Consistent with practical UAV-assisted cellular and RIS-enabled deployments, dedicating each UAV-mounted RIS to a single GRU cluster enables stable control signaling, efficient interference management, and predictable resource coordination~\cite{8736350}. 

In the considered UAV-mounted RIS-assisted TN-NTN system, the locations of HAP, GRUs, UAV-mounted RISs, and IIoT devices at time slot $t$ are denoted by $\mathbf{Q}_h=(x_h,y_h,z_h)$, $\mathbf{Q}_g^t=(x_g^t,y_g^t,Z_g)$, $\mathbf{Q}_u^t=(x_u^t,y_u^t,Z_u)$, and $\mathbf{Q}_i=(x_i,y_i,0)$, respectively. The service horizon $T$ is divided into $N$ discrete time slots indexed by $t \in \mathcal{T}=\{1,2,\ldots,N\}$, each with duration $\delta=T/N$, over which network control actions are executed at different timescales. 
Throughout the operation, UAV-mounted RISs passively assist GRU transmissions by sharing the available spectrum and adaptively reflecting signals toward intended IIoT devices, while coordinated interference management across aerial and terrestrial links ensures reliable and spectrally efficient industrial communications.


\subsection{Wireless Communication Model}
This section presents the channel model of the proposed framework. we consider both the direct terrestrial link between GRUs and IIoT devices, HAP and IioT devices, and the indirect RIS-assisted link involving UAV-mounted RISs. All transmissions operate over the OFDMA, and the channel depends on distance, and propagation environment. Let us define the access association as $\alpha_{i,j,x}^t \in \{0,1\}$, where $x \in \{g,u,h\}$. When $\alpha_{i,j,g}^t=1$, it means that the IIoT $i$ is connected with GRU $g$ to execute task $j$ at time slot $t$.

\subsubsection{GRU-IIoT communication}

The GRU allocates subchannels to IIoT devices to support the transmission of industrial data at time slot $t$.
Let us denote the distance between GRU $g$ and device $i$ as $d_{i,j,g}^t=\sqrt{(x_g-x_i)^2+(y_g-y_i)^2+Z_g^2}$. The corresponding large-scale path loss at time $t$ is $PL_{i,j,g}^t=\psi_{\text{LoS}}\!\left(\frac{4\pi f_c d_{i,j,g}^t}{c}\right)^{\!2}
+\psi_{\text{NLoS}}\!\left(\frac{4\pi f_c d_{i,j,g}^t}{c}\right)^{\!2},$ and the instantaneous channel gain is $\eta_{i,j,g}^t = \bar{g}_{i,j,g}^t |\widetilde{\eta}_{i,j,g}^t|^2,
\quad \bar{g}_{i,j,g}^t = \frac{1}{PL_{i,j,g}^t},\;
|\widetilde{\eta}_{i,j,g}^t|^2 \sim \mathrm{Exp}(1).$ Hence, The SINR between IIoT device $i$ and GRU $g$ in service slice 
$j$ is expressed as, 
$\gamma^t_{i,j,g}= \frac{\alpha_{i,j,g}^t|\eta_{i,j,g}^t|^2 P_{i,j,g}^t}{\sum_{k \in \bar{N}: k \neq i} \alpha_{k,j,g}^t |\eta_{k,j,g}^t|^2 P_{k,j,g}^t + \sigma^2}$,
The achievable data rate can be expressed as:
    $R_{i,j,g}^t = B_{i,j,g}^t \log_2(1+\gamma^{t}_{i,j,g})$
Moreover, the transmission latency and energy consumption can be calculated as: $L_{i,j,g}^t=\frac{D_{i,j}}{R_{i,j,g}^t}, E_{i,j,g}^t=\alpha_{i,j,g}^tP_{i,j,g}^tL_{i,j,g}^t$, respectively.


\subsubsection{HAP-IIoT communication}
When the GRUs face a congestion and cannot process the received data, the data is transferred to the higher layer of NTN. Let $d_{i,j,h}^t$ denote the distance between HAP $h$ and IIoT device $i$. Due to obstacles and industrial structures, the link may experience either LoS or NLoS propagation. Hence, we formulate the model using Rician fading model. The path loss is
$PL_{i,j,h}^t =
P_{i,j,h}^{LoS,t} \psi_{\mathrm{LoS}}\!\left(\frac{4\pi f_c d_{i,j,h}^t}{c}\right)^2
+ P_{i,j,h}^{\mathrm{NLoS,t}} \psi_{\mathrm{NLoS}}\!\left(\frac{4\pi f_c d_{i,j,h}^t}{c}\right)^2$, where $P_{i,j,h}^{LoS,t}$ is the LoS probability and can be calculated as $P_{i,j,h}^{LoS,t} = \frac{1}{1+\omega_1 \exp{(-\omega_2 [\theta_{i,j,h}^t-\omega_1])}}$, $P_{i,j,h}^{NLoS,t} = 1-P_{i,j,h}^{LoS,t}$, $\theta_{i,j,h}^t = \frac{180}{\pi} \times \sin^{-1}(\frac{z_h}{d_{i,j,h}^t})$, and the instantaneous channel gain is
$\eta_{i,j,h}^t = \bar{g}_{i,j,h}^t |\widetilde{\eta}_{i,j,h}^t|^2,
\bar{g}_{i,j,h}^t = \frac{1}{PL_{i,j,h}^t},$
and $|\widetilde{\eta}_{i,j,h}^t|^2
= \sqrt{\frac{K}{K+1}}\psi_{\mathrm{LoS}}
+ \sqrt{\frac{1}{K+1}}\psi_{\mathrm{NLoS}},$
where \(K\) is the Rician factor.

The SINR between IIoT device $i$ and HAP $h$ in service slice 
$m$ is expressed as, 
$\gamma^{t}_{i,j,h}= \frac{\alpha_{i,j,h}^t|\eta_{i,j,h}^t|^2 P_{i,j,h}^t}{\sum_{k \in \bar{N}: k \neq i} \alpha_{k,m,h}^t |\eta_{k,m,h}^t|^2 P_{k,m,h}^t + \sigma^2}$. Hence, the achievable data rate can be expressed as:
    $R_{i,j,h}^t = B_{i,j,h}^t \log_2(1+\gamma^{t}_{i,j,h})$
Moreover, the transmission latency and energy consumption can be calculated as $L_{i,j,h}^t=\frac{D_{i,j}}{R_{i,j,h}^t}, E_{i,j,h}^t=\alpha_{i,j,h}^tP_{i,j,h}^tL_{i,j,h}^t$, respectively.


\subsubsection{UAV-mounted RIS-IIoT communication}
The indirect link is through the UAVs-mounted RIS to IIoT devices. Each UAV is equipped with an RIS of $M$ passive reflecting elements arranged in a uniform linear array. The indirect path consists of the HAP-UAV-mounted RIS followed by the UAV-mounted RIS-IIoT, both modeled by Rician fading due to dominant LoS components.

We assume that the RIS elements are arranged in a linear array, and the communication links between the IoT devices and UAV-RIS, as well as between UAV-RIS and the central base station \( B_c \), are subject to Rician fading. Each UAV-RIS \( u \) is equipped with \( M \) phase shift elements, where $m \in \Bar{M} = \{1,\cdots, M\}$. 
The communication link between IIoT device \( i \) and UAV-RIS \( u \), denoted as \( \eta_{i,j,u}^t \), is modeled as:
    $\eta_{i,j,u}^t = \sqrt{\frac{\varrho}{(d^t_{i,u})^2}} \times \sqrt{\frac{Z_{Rician}}{Z_{Rician}+1}} \times \widetilde{\eta}_{i,j,u}^t
    \label{channel_gain}$.
where $d^t_{i,u}= \sqrt{(x_i^t - x_u^t)^2 +(y_i^t - y_u^t)^2 + (h_u^t)^2}$ is the distance between IoT device $i$ and UAV-RIS $u$, $Z_{Rician}$ denotes the Rician parameter. Also, $\varrho$ is the average path loss power gain at reference distance $d_0=1$ m. $\widetilde{\eta}_{i,j,u}^t$ is the LoS array elements of $i$ and $u$ and is expressed as:
\begin{equation}
    \widetilde{\eta}_{i,j,u}^t = \underbrace{\Biggr[1, e^{-j\frac{2\pi}{\lambda}\tau\phi_{i,u}^t}, \dots , e^{-j\frac{2\pi}{\lambda}(M-1)\tau\phi_{i,u}^t} \Biggr]^T}_{\text{array components}}
    \label{Array_elements}
\end{equation}
where $\phi_{i,u}^t = {(x_i^t - x_u^t)}/{d_{i,u}^t}$ is the cosine of the angle of the signal traversing from the UAV $u$ to the IoT $i$, $\tau$ represents the RIS elements separation, and $\lambda$ is the wavelength of the carrier signal.
Likewise, the channel gain between UAV $u$ and HAP $h$, denoted by $\eta_{h,j,u}^t \in \mathbb{\eta}^{M \times 1}$, can be calculated similarly to Eq.~\eqref{Array_elements}, where $\widetilde{\eta}_{h,j,u}^t$ being the LoS array elements of UAV $u$ and HAP $h$ with $\phi_{h,u}^t = {(x_u^t - x_{h}^t)}/{d_{h,u}^t}$ being the cosine of the angle of the signal traversing from the UAV-RIS $u$ to HAP $h$.

The phase shift matrix of UAV-RIS $u$ during the $t$-th time slot can be expressed as $\Theta_u^t = \text{diag}\{e^{j\Phi_{1,u}^t}, \dots, e^{j\Phi_{M,u}^t}\}$, with $\Phi^t_{m,u} \in \{0, {2\pi}/{W}, \dots, {2\pi (W-1)}/{W}\} = \psi$ representing the phase shift of the $m$-th array component at time slot $t$. Here, $W = 2^g$ with $g$ being the number of bits identifying the RIS phase elements array components. 
Consequently, the SINR $\gamma^{t}_{i,j,u}$ of the signal received by the base station $B_c$ through UAV-RIS $u$ and sent by IoT device $i$ is: 
$$\gamma^{t}_{i,j,u}= \frac{\alpha_{i,j,u}^t|\eta_{i,j,u}^{t,H}\Theta_u^t \eta_{h,j,u}^t|^2 P_{i,j,u}^t}{\sum_{k \in \bar{N}: k \neq i} \alpha_{k,j,v}^t |\eta_{k,v}^{t,H}\Theta_v^t \eta^t_{B_c,v}|^2 P_k +\sigma^2},$$
where $\sigma^2$ is the Gaussian noise power, $H$ represents the operator of the conjugate transpose, $k$ is the index number of interfering IoT devices transmitting data to the UAVs at time $t$, $P_i$ and $P_k$ is the transmit power of the IoT device $i$ and the interfering IoT devices $k$, respectively.
%

The data rate of the data transferred from IoT device $i$ to base station $B_c$ through UAV-RIS $u$ can be expressed as:
$R_{i,j,u}^t = B_{i,j,u}^t \log_2(1+\gamma^{t}_{i,j,u})$
where $B_{i,j,u}^t$ is the allocated bandwidth. Moreover, the transmission latency and energy consumption can be calculated as $L_{i,j,u}^t=\frac{D_{i,j}}{R_{i,j,u}^t}, E_{i,j,u}^t=\alpha_{i,j,u}^t P_{i,j,u}^tL_{i,j,u}^t$, respectively.
Moreover, the total data rate $R_{i,j}^{t,tot}$ to transmit data from IIoT $i$ to the computing node to execute the task $j$ can be given by:
$R_{i,j}^{t,tot} = \alpha_{i,j,g}^t R_{i,j,g}^t + \alpha_{i,j,h}^t R_{i,j,h}^t + \alpha_{i,j,u}^t R_{i,j,u}^t$.

\subsection{Energy Consumptions Model}

Let $K_{i,j}$ denote the size of industrial computing tasks generated by IIoT device $i$. Each task is represented as $K_{i,j}=\{D_{i,j},\,F_{i,j}^{req},\,L_{i,j}^{max}\}$
where $D_{i,j}$ is the input data size of task $j$ from IIoT $i$, $F_{i,j}^{req}$ denotes the required CPU cycles per bit, and $L_{i,j}^{max}$ is the maximum latency required to compute the task $j$ that is related to IIoT $i$. 
For each task $D_{i,k}$, an offloading ratio $\Psi_{i,j}^t\in[0,1]$ is adopted, where $\Psi_{i,j}^t=0$ indicates local execution, $\Psi_{i,j}^t=1$ denotes full offloading, and $0<\Psi_{i,j}^t<1$ represents partial offloading. Hence the local latency for executing the task $j$ at IIoT $i$ can be expressed as $L^{t,loc}_{i,j} = \frac{(1-\Psi_{i,j}^t))D_{i,j}\,F_{i,j}^{req}}{F_i},$ where $F_i$ denotes the computing capability of IIoT device $i$. The local energy consumption can be expressed as $E_{i,j}^{t,loc} = D_{i,j}F_{i,j}^{req}F_i^3L^{t,loc}_{i,j}$


If a fraction $\Psi_{i,j}^t$ is offloaded, it is transmitted to a MEC server deployed either at the GRU or at the HAP (directly or via the UAV-mounted RIS). The energy consumption of executing task $j$ at the MEC server located at GRU $g$ can be written as: $E_{j,g}^{t,exec} = \sum_{i \in \mathcal{I}} \alpha_{i,j,g}^t (F_{i,j,g}^t)^3L_{i,j,g}^{t,exec},$ where $F_{i,j,g}^t$ is a constant specifying the chip architecture of the GRU and its capability of processing the received industrial tasks, and $L_{i,j,g}^{t,exec}$ is the latency of the latency of executing task $j$ coming from IIoT $i$ at GRU $g$ and $L_{i,j,g}^{t,exec} = \frac{D_{i,j}\,F_{i,j}^{req}}{F_{i,j,g}^t}$. Similarly, the energy consumption of executing task $j$ at the MEC server located at HAP $h$ can be written as: $E_{j,h}^{t,exec} = \sum_{i \in \mathcal{I}} \alpha_{i,j,h}^t (F_{i,j,h}^t)^3L_{i,j,g}^{t,exec},$ where $F_{i,j,h}^t$ is a constant specifying the chip architecture of the GRU and its capability of processing the received industrial tasks, and $L_{i,j,h}^{t,exec}$ is the latency of the latency of executing task $j$ coming from IIoT $i$ at HAP $h$ and $L_{i,j,h}^{t,exec} = \frac{D_{i,j}\,F_{i,j}^{req}}{F_{i,j,h}^t}$.

The overall latency and energy consumption can be expressed as:
    $L_{i,j}^{t} = L_{i,j,g}^t + L_{i,j,h}^t + L_{i,j,u}^t + L_{i,j,g}^{t,exec} + L_{i,j,h}^{t,exec}.$
Hence, the overall computing latency can be given as:
    $L_{i,j}^{t,tot} = \max{\{L_{i,j}^{t,loc}, L_{i,j}^{t}\}}.$
The overall energy consumption which include the execution and transmission energy can can be given by:
    $E_{i,j}^{t,tot} = E_{i,j,g}^t + E_{i,j,h}^t + E_{i,j,u}^t + E_{i,j,g}^{t,exec} + E_{i,j,h}^{t,exec}.$
\section{Problem Formulation}
\label{problem_formulation}
As industrial IIoT applications, stringent URLLC requirements, together with energy-constrained devices and the need for high-throughput data exchange, make the simultaneous minimization of latency and energy consumption and maximization the data rates essential. To capture these conflicting objectives within a unified framework, we formulate the objective function that integrates communication efficiency and system cost \cite{10145918}. Specifically, the transmission performance of IIoT device $i$ at time slot $t$ is quantified by the logarithmic utility terms where $\psi^{t}= \ln\!\big(1+R^{t,tot}_{i,j}\big)$. 
This rate-utility component characterizes the communication benefit of IIoT transmissions and are incorporated into the numerator, while latency and energy consumption are included as penalty terms in the denominator, thereby enabling the final optimization problem to simultaneously promote high data rates and low operational cost. Hence, the utility function to be maximized can be given by:
\begin{equation}
U^{t}_{i,j} =
\frac{\psi^{t}}{\omega_t\,(L_{i,j}^{\mathrm{t,tot}}/L_0)+\omega_e\,(E_{i,j}^{\mathrm{t,tot}}/E_0)
}
\end{equation}
where $\omega_l, \text{ and } \omega_e$ are latency, energy weights, $L_0, \text{ and } E_0$ are latency and energy normalization factors.

The aim of the proposed optimization is to jointly determine how IIoT devices offload their tasks and how communication and computing resources are distributed across the terrestrial and non-terrestrial layers. This includes the coordinated allocation of and downlink spectrum, transmission power, MEC computing capacity at GRUs and HAPs, and the configuration of UAV-mounted RIS, while ensuring an efficient association between IIoT devices and their serving computing nodes. Hence, the overall optimization problem can be written as:
\begin{normalsize}
\begin{flalign}
    &\max_{\mathcal{F},
    \alpha,
    \Psi,
    \Theta,
    \mathcal{B},\mathcal{P}} 
    \sum_{t \in \mathcal{T}}
    \sum_{i \in \mathcal{I}}
    \sum_{j \in \mathcal{J}}
    U^{t}_{i,j} 
    \label{objective_fun}\\
    & \mathrm{subject~to:}\nonumber\\
    &C_1:\alpha_{i,j,x}^t \in \{0,1\}, \forall t \in \mathcal{T},\forall i \in \mathcal{I}, \forall j \in \mathcal{J}, \forall x \in \{\mathcal{G},\mathcal{U},H\}
    \tag{\ref{objective_fun}a}
         \label{C1}\\
    & C2: \Psi_{i,j}^t\in[0,1], \forall t \in \mathcal{T},\forall i \in \mathcal{I}, \forall j \in \mathcal{J},
         \label{C2}
         \tag{\ref{objective_fun}b}\\
    & C3: \sum_{g \in \mathcal{G}}\alpha_{i,j,g}^t+
    \alpha_{i,j,h}^t+
    \sum_{u \in \mathcal{U}}\alpha_{i,j,u}^t \leq 1,
    \label{C3}
         \tag{\ref{objective_fun}c}\\
    & C4: 0 \leq \sum_{x=1}^{X} B_{i,j,x}^t \leq B^{max}_x,\forall i \in \mathcal{I}, \forall j \in \mathcal{J}
    \label{C4}
    \tag{\ref{objective_fun}d}\\
    & C5: 0 \leq \sum_{x=1}^{X}P_{i,j,x}^t \leq P^{max}_x,\forall i \in \mathcal{I}, \forall j \in \mathcal{J},
     \label{C5}
     \tag{\ref{objective_fun}e}\\
     & C6:  0 \leq \sum_{x=1}^{X}F_{i,j,x}^t \leq F^{max}_x,\forall i \in \mathcal{I}, \forall j \in \mathcal{J},
     \label{C6}
     \tag{\ref{objective_fun}f}\\
    &  C7: 0 \leq \Phi_{j,u}^t \leq 2\pi,\forall t \in \mathcal{T},\forall i \in \mathcal{I}, \forall j \in \mathcal{J},
    \label{C7}
    \tag{\ref{objective_fun}g}\\
    &  C8: R_{i,j}^{t,tot} \geq R^{min},\forall t \in \mathcal{T},\forall i \in \mathcal{I}, \forall j \in \mathcal{J},
    \label{C8}
    \tag{\ref{objective_fun}h}\\
    & C9:L_{i,j}^{t,tot} \leq L^{max}, \forall t \in \mathcal{T},\forall i \in \mathcal{I}, \forall j \in \mathcal{J},
    \label{C9}
    \tag{\ref{objective_fun}i}\\
    & C10: E_{i,j}^{t,tot} \leq E^{max}, \forall t \in \mathcal{T},\forall i \in \mathcal{I}, \forall j \in \mathcal{J},
    \label{C10}
    \tag{\ref{objective_fun}j}
\end{flalign}
\end{normalsize}

In the proposed formulation, $\alpha=\alpha_{i,j,x}^t\in\{0,1\}$ denotes the binary association variable indicating whether IIoT device $i$ is associated with serving entity $x\in\{\text{GRU},\text{UAV},\text{HAP}\}$ at time slot $t$. $\mathcal{F}=\{F^t_{i,,j,x}\}, \mathcal{B}=\{B^t_{i,,j,x}\}, \mathcal{P}=\{P^t_{i,,j,x}\}$ represent the allocated resources, bandwidth, and transmission power, respectively. $\Theta = \Theta_u^t$ represents the RIS phase shift elements, and $\Psi = \Psi_{i,j}^t$ represent the offloading decision ratio.
The constraints in (\ref{objective_fun}) are defined as follows. Constraint C1 enforces binary association variables, ensuring that user-node association decisions are discrete. Constraint C2 bounds the normalized resource utilization variables within feasible limits. Constraint C3 guarantees that each user is associated with at most one serving entity at any time slot, preventing conflicting access across terrestrial and non-terrestrial layers. Constraints C4 and C5 limit the total allocated bandwidth and transmit power, respectively, according to the maximum capabilities of each serving node. Constraint C6 restricts the assigned computation resources to the available processing capacity. Constraint C7 confines the RIS phase shifts within $[0,2\pi]$, reflecting hardware limitations. Constraint C8 ensures that the achieved and downlink data rates satisfy minimum QoS requirements. Constraint C9 imposes an upper bound on the end-to-end latency, while Constraint C10 limits the total energy consumption to guarantee energy-efficient operation.

\section{The Proposed MADRL-based solutions}
\label{propsed_solution}
The optimization problem formulated in (\ref{objective_fun}) delineates the joint control of multiple continuous resource allocation variables and RIS configurations and discrete association decisions across heterogeneous radio and computing entities. The high dimensionality of the state-action space, coupled with strongly interdependent QoS, latency, and energy objectives, renders conventional centralized optimization computationally prohibitive and obtained solutions are intractable. To address this challenge, the problem is first modeled as a stochastic decision-making process that captures the dynamic interactions between IIoTs, radio resources, and RIS configurations. Subsequently, a distributed multi-agent learning framework is adopted to enable efficient and scalable resource allocation and computations, while preserving adaptability to time-varying network conditions.
\subsection{Dec-POMDP}
The proposed resource management problem is modeled as a Dec-POMDP to capture the distributed decision-making nature of the Open-RAN-enabled TN/NTN architecture. According to our framework, the agents including GRUs, HAP and UAV-mounted RIS controllers operate concurrently and make decisions based on partial and local observations, while jointly contributing to a common system objective. The Dec-POMDP is defined by the tuple
$\{ \mathcal{K}, \mathcal{S}, \{\mathcal{A}_k\}_{k\in\mathcal{K}}, \mathcal{P}, \mathcal{R}, \{\mathcal{O}_k\}_{k\in\mathcal{K}} \}$,
where $\mathcal{K}$ denotes the set of agents, $\mathcal{S}$ is the global state space, $\mathcal{A}_k$ and $\mathcal{O}_k$ represent the action and observation spaces of agent $k$, respectively, $\mathcal{P}$ is the state transition probability, and $\mathcal{R}$ is the shared reward function.
\subsubsection{State}  
The state set aim to capture the network-wide information, including GRU and UAV-mounted RIS locations, channel conditions, traffic demands, residual energy levels, and latency indicators. 
Specifically, $O_k^t = \{\gamma^t_{i,j,x},d_{i,j,x}^t,D_{i,j}^t\}$.

\subsubsection{Action Space}  
At each time slot $t$, agents select actions from heterogeneous continuous action spaces whose dimensions depend on the agent type. Specifically, GRU agents determine user association variables $\alpha_{i,j,g}^t$, bandwidth allocation $\mathcal{B}_{i,j,g}^t$, transmission power $\mathcal{P}_{i,j,g}^t$, and local computing resource allocation $\mathcal{F}_{i,j,g}^t$. HAP agents similarly control communication and computing resources for their associated IIoT devices at the aerial MEC layer. In contrast, UAV-mounted RIS agents operate over a lower-dimensional action space that exclusively adjusts the RIS phase-shift configuration $\Theta_u^t$. 


\subsubsection{Reward Function}  
The reward is designed to promote cooperative behavior among agents by maximizing the aggregate QoS of IIoT devices, while penalizing excessive latency, energy consumption, and QoS violations. This shared reward encourages agents to coordinate implicitly and achieve system-wide performance objectives despite operating with partial observability. Specifically, $R^t = \sum_{i \in \mathcal{I}}\sum_{j \in \mathcal{J}} U_{i,j}^t$.


To solve the Dec-POMDP defined in the previous subsection, we adopt a multi-agent proximal policy optimization (MAPPO) algorithm. Let $\boldsymbol{\pi}=\{\pi_k\}_{k\in\mathcal{K}}$ denote the set of agent policies, where each agent selects actions based solely on its local observation during execution.
During training, a centralized critic $V(s;\omega)$ exploits the global state information to estimate the joint value function, thereby mitigating the non-stationarity induced by decentralized learning. Each agent’s actor network $\pi_k(a_k|o_k;\theta_k)$ is updated using the PPO clipped surrogate objective with advantages computed from the centralized critic. The critic parameters are optimized by minimizing the mean squared error between the predicted value and the empirical return.
This MAPPO framework enables cooperative policy learning among distributed agents while preserving decentralized execution, making it well suited for large-scale, partially observable wireless systems. Algorithm ~\ref{alg:mappo} delineates the steps followed by MAPPO to achieve the convergence and learning the optimal policies.

\begin{algorithm}[t]
\caption{MADRL-based PPO Learning Framework}
\label{alg:mappo}
\begin{algorithmic}[1]
\State \textbf{Initialize:} actor policies $\{\pi_k(\cdot;\theta_k)\}_{k\in\mathcal{K}}$, centralized critic $V(\cdot;\omega)$, replay buffer $\mathcal{D}$
\For{each episode $e=1$ to $Episodes$}
    \State Reset environment and observe initial global state $s_0$
    \For{each time slot $t=1$ to $T$}
        \For{each agent $k \in \mathcal{K}$}
            \State Observe local observation $o_k^t$
            \State Select action $a_k^t \sim \pi_k(\cdot|o_k^t;\theta_k)$
            \If{~\ref{C4},~\ref{C5},~\ref{C6},~\ref{C8},~\ref{C9},~\ref{C10}}
            \State $R^t += U_{i,j}^t$
            \ElsIf{any disrespect of constraints of ~\ref{C4},~\ref{C5},~\ref{C6},~\ref{C8},~\ref{C9},~\ref{C10}}
            \State $R^t -= Constant$
            \EndIf
        \EndFor
        \State Execute joint action $\mathbf{a}^t=\{a_k^t\}_{k\in\mathcal{K}}$
        \State Observe reward $r_t$, next state $s_{t+1}$
        \State Store $(s_t,\mathbf{a}^t,r_t,s_{t+1})$ in $\mathcal{D}$
    \EndFor
    \State Compute advantages $A_t=r_t+\gamma V(s_{t+1})-V(s_t)$
    \For{each agent $k \in \mathcal{K}$}
        \State Update actor $\theta_k$ using PPO clipped objective
    \EndFor
    \State Update critic $\omega$ by minimizing value loss
\EndFor
\end{algorithmic}
\end{algorithm}

\section{simulation results and analysis}
\label{Performance_evaluation}
The performance of the proposed MAPPO-based framework is evaluated through comprehensive simulations and compared against three benchmark schemes: MAPPO without RIS assistance, MAAC approach, and a greedy baseline, which is based on selecting the best actions at that time step. These comparisons are conducted to assess the effectiveness of the proposed solution in jointly optimizing computing resources, spectrum sharing, transmission power, bandwidth allocation, and RIS phase configurations. The simulation environment covers a $1000m \times 1000m$  industrial area with 100 IIoT organized into three serving clusters, in which each IIoT generate three different tasks at the same time step. To capture realistic industrial deployment characteristics, the spatial distribution of GUs follows a power-law model, reflecting nonuniform and hotspot-driven IIoT densities. All experiments are implemented using TensorFlow 1.13.1 and Python 3.7.16 on a Windows platform equipped with an Intel Xeon i7 processor (2.2 GHz, 4 cores) and 16 GB RAM. The key system parameters are set as follows: noise power $\sigma^2=-95dBm$, packet size $K_i=100 MB$, discount factor $\gamma=0.99$, and carrier frequency $f_c=5$ GHz. Furthermore, the maximum allowable latency, bandwidth, and transmission power are constrained to $L_{max}=20ms, B_{max}=100MHz, \text{ and } P_{max}=35dBm$, respectively, enabling a realistic and dynamic performance evaluation of the proposed framework. 

\subsection{Convergence Analysis}
Fig.~\ref{latency} compares the average end-to-end latency of different schemes as the number of IIoT devices increases. The proposed MAPPO framework with RIS consistently achieves the lowest latency across all network loads. For instance, at 160 IIoT devices, MAPPO with RIS reduces latency, yielding a 17\% reduction compared to the MAAC scheme and a 20\% improvement over MAPPO without RIS. Compared with the greedy baseline, the proposed approach achieves up to a 35\% latency reduction. These results confirm the advantage of combining RIS-enabled propagation control with cooperative multi-agent learning for latency-critical industrial 6G applications.

\subsection{Data Rate Analysis}
\begin{figure}[t!]
\centering
	\mbox{
	    \hspace{-7mm} 
        \subfigure[\label{convergence}]{\includegraphics[scale=0.25]{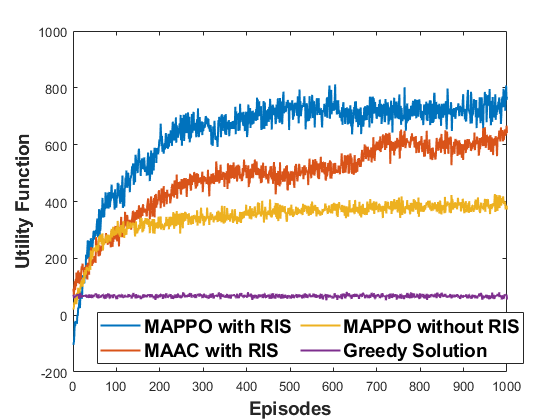}
	   }
	     \subfigure[\label{data_rates}]{\includegraphics[scale=0.25]{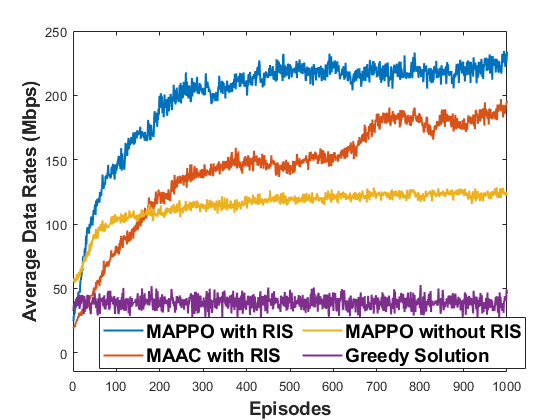}}
} 
	\caption{Proposed solution comparison with benchmark algorithms in terms of convergence and data rate.}
	\label{fig:graph4}
\end{figure}

Fig.~\ref{data_rates} illustrates the convergence of the average data rate achieved by the proposed MAPPO framework with RIS assistance compared to MAPPO without RIS, MAAC, and a greedy baseline. The proposed MAPPO with RIS converges rapidly and achieves the highest steady-state data rate of approximately 220 Mbps. This corresponds to an improvement of about 25\% over the MAAc scheme and nearly 75\% compared to MAPPO without RIS. In contrast, the greedy solution remains limited, resulting in lower throughput than the proposed approach. These results demonstrate that the integration of RIS significantly enhances spectral efficiency and data-rate performance. 

\subsection{Latency Analysis}
Fig.~\ref{latency} compares the average end-to-end latency of different schemes as the number of IIoT devices increases. The proposed MAPPO framework with RIS consistently achieves the lowest latency across all network loads, yielding a 20\% reduction compared to the MAAC scheme and a 25\% improvement over MAPPO without RIS. Compared with the greedy baseline, the proposed approach achieves up to a 32\% latency reduction. These results confirm the advantage of combining RIS-enabled propagation control for latency-critical industrial 6G applications.

\subsection{Energy Consumption Analysis}
\begin{figure}[t!]
\centering
	\mbox{
	    \hspace{-7mm} 
        \subfigure[\label{latency}]{\includegraphics[scale=0.15]{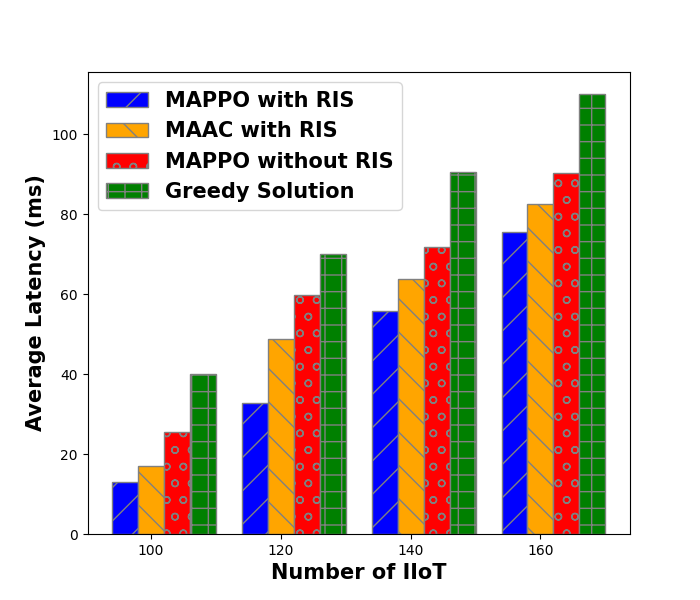}
	   }
	     \subfigure[\label{Energy_consumption}]{\includegraphics[scale=0.15]{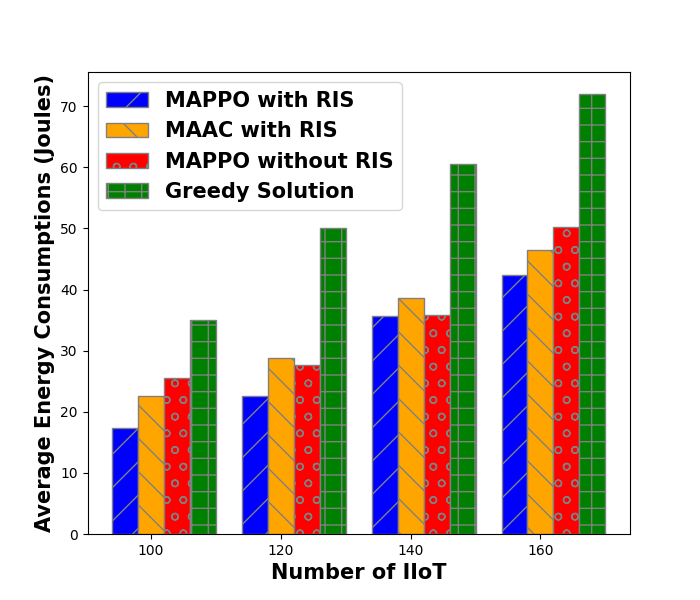}}
} 
	\caption{Proposed solution comparison with benchmark algorithms in terms of latency and energy consumption when increasing the IIoTs}
	\label{fig:graph5}
\end{figure}

Fig.~\ref{Energy_consumption} compares the average energy consumption of different schemes as the number of IIoT devices increases. The proposed MAPPO framework with RIS consistently achieves the lowest energy consumption across all traffic loads. For example, at 160 IIoT devices, MAPPO with RIS reduces energy consumption, achieving about a 12\% reduction compared to the MAAc approach and nearly a 20\% improvement over MAPPO without RIS. Compared with the greedy baseline, the proposed method yields an energy saving of up to 40\%, highlighting its superior energy efficiency, demonstrating that RIS-assisted MAPPO effectively mitigates energy growth under increasing network load.
\section{conclusion}
\label{conclusion}
In this work, we investigated QoS-aware resource optimization in RIS-enabled Open-RAN industrial 6G networks with integrated TN/NTN layers. By exploiting UAV-mounted RISs in cooperation with IIoTs and a HAP, a decentralized MAPPO-based framework was developed to jointly optimize spectrum sharing, power allocation, computing resources, and RIS configurations under dynamic industrial conditions. Simulation results demonstrate that the proposed approach significantly improves data rates, reduces latency, and lowers energy consumption compared with state-of-the-art benchmarks, highlighting the potential of RIS-assisted Open-RAN intelligence for scalable and energy-efficient industrial 6G networks.
Future work will extend the proposed framework to jointly consider dynamic UAV mobility, downlink transmission, and federated or transfer learning mechanisms to enhance scalability, robustness, and real-time adaptability in large-scale industrial 6G deployments.
\section*{Acknowledgment}
This work was supported in whole, or in part, by the Luxembourg National Research Fund (FNR), ref. C22/IS/17220888/RUTINE.
\bibliographystyle{IEEEtran}
\bibliography{bibliography.bib}
\end{document}